\documentclass[twocolumn,aps,prl,showpacs,groupedaddress]{revtex4}
\usepackage{epsfig,amssymb,amsmath}
\usepackage[hypertex,linkcolor=red]{hyperref}
\def\comment#1{}\def\labell#1{\label{#1}}
\def\Cmplx{\mathbb C}

\def\spc#1{\mathcal{#1}}
\def\SU#1{\mathbb{SU}(#1)}
\def\<{\langle}\def\>{\rangle}
\def\d{\operatorname{d}}

\begin{document}
\title{Secret quantum communication of a reference frame}
\author{Giulio Chiribella, Lorenzo Maccone, Paolo
  Perinotti}\affiliation{QUIT - Quantum Information Theory Group,
  Dipartimento di Fisica ``A.  Volta'' Universit\`a di Pavia, via A.
  Bassi 6, I-27100 Pavia, Italy.}
\date{\today}

\begin{abstract}
  We propose quantum cryptographic protocols to secretly communicate a
  reference frame---\emph{unspeakable} information in the sense it
  cannot be encoded into a string of bits. Two distant parties can
  secretly align their Cartesian axes by exchanging $N$ spin 1/2
  particles, achieving the optimal accuracy $1/N$. A possible
  eavesdropper cannot gain any information without being detected.
\end{abstract}
 \maketitle

Any communication about the properties of a physical object requires
not only the transmission of a string of bits, but also the existence
of notions shared between sender and receiver. For example, to
communicate the length of an object, they must agree on a unit of
length. In many cases, the natural units of length, time, and
mass~\cite{planck} provide a common standard that allows any two
parties to communicate by simply exchanging bits, even though they
have never met.  However, such standard references do not exist for
all properties, as some cannot be communicated as strings of
bits~\cite{gisin,refframe} and are hence referred to as
\emph{unspeakable information}~\cite{petra,popescu}.  For example, due
to the fundamental isotropy of space, a string of bits can encode only
the \emph{relative} orientation of two directions, but not the
\emph{absolute} orientation of a single direction in space. In such
cases, the only possibility for the sender to establish a shared
notion with the receiver is to send a physical object---such as a
gyroscope---that provides the common standard.  This situation arises
whenever sender and receiver try to establish a shared reference frame
(SRF), e.g. by aligning their Cartesian axes, or by synchronizing
their clocks.

Using quantum cryptography, one can securely send speakable
information (i.e. strings of bits) through a public channel. One might
then ask whether the same is true for unspeakable information. The
answer is not trivial since to convey unspeakable info a physical
carrier must be sent, and it can be intercepted by an eavesdropper. In
this paper we present quantum cryptographic protocols to encode
spatial directions and reference frames so that no eavesdropper can
obtain information on them. This is useful also for quantum
cryptography, since private SRFs are an important resource to realize
both classical and quantum secret communication~\cite{qcrypt}.  Here,
we consider communication through the transmission of $N$ spin 1/2
particles. We start by showing a simple, experimentally feasible
protocol that uses $N$ shared secret bits and $N$ separable states to
communicate a secret direction with accuracy $1/\sqrt{N}$. Then, we
present the optimal protocol, which exploits entanglement and secretly
transmits a whole Cartesian frame. Such a protocol employs $3 \log N$
secret bits to achieve accuracy $1/N$.  Remarkably, this amount of
secret bits exactly coincides with the number of secret bits that can
be exchanged once a private SRF is established~\cite{qcrypt}. Such a
tight balance provides a new understanding of reference frames as a
communication resource. Note that the above protocols do not need the
two parties to possess the shared secret bits beforehand: They can
obtain them using quantum key distribution, which does not necessarily
require a prior shared reference~\cite{prl,altri}. This suggests that
protocols that do not use prior classical randomness can be
implemented. We conclude the paper by providing two such protocols
that require only quantum and classical communication over a public
channel.

{\em Separable protocol using secret random bits.}---The intuitive
idea of this protocol is that two parties (say Alice and Bob) can
transform a secret shared random string into a secret shared
direction.  Imagine that Alice wants to communicate to Bob the secret
direction of her $z$-axis. For each 0 and 1 in the secret string, she
sends a spin 1/2 particle pointing respectively up and down (according
to her own $z$ axis). Then, Bob just needs to rotate the $z$ axis of
his Stern-Gerlach apparatus until its measurement results (0 for
``up'' and 1 for ``down'') match the bits of the shared secret string.
When this happens, he knows that he has aligned his $z$ axis with
Alice's.  A possible eavesdropper Eve will not be able to gain any
information on the direction of the $z$ axis, since she does not know
the secret string. In fact, from her point of view a random sequence
of spin up/spin down is equivalent to the maximally chaotic state,
where the spins are randomly oriented {\em in any direction} she might
choose to sample. Moreover, if Eve is tampering with the
communication, Bob will find it out as any direction he chooses for
his Stern-Gerlach apparatus will never yield the secret string as
outcome bits.

A more practical implementation of the above protocol can be obtained
if Bob performs all measurements aligning his Stern-Gerlach apparatus
alternately along his own $x$, $y$ and $z$ axes. The probability that
Alice's spin up states (identified by zeros in the shared secret key)
will have outcome ``up'' in Bob's $z$-oriented Stern-Gerlach apparatus
is $p(\theta)=\cos^2\theta/2$, where $\theta$ is the unknown angle
between Alice and Bob's $z$ axes. This is also the probability that
Alice's spin down states (identified by ones) will have outcome
``down'' at Bob's apparatus. Therefore, Bob can estimate such
probability from his outcomes on $N$ spins as $\bar
p_\theta=[N($up$,0)+N($down$,1)]/N$, where $N(i,j)$ is the number of
spins that gave outcome $i$ when the corresponding shared secret bit
was $j$.  Once the angle $\theta$ has been estimated from $\bar
p_\theta$, Bob orients his Stern-Gerlach apparatus along his $x$ and
$y$ axes, and repeats the procedure to recover the angles between
these and Alice's $z$ axis.  The eavesdropper Eve can be detected by
Bob. In fact, her action would result in a depolarization of the
transmitted spins and the three angles Bob recovers would be
inconsistent. He can discover it if the sum of their squared cosines
differs from one by more than a purely statistical error would allow.
The accuracy of Bob's procedure can be evaluated as follows.  The rms
error in the estimate of $p(\theta)$ from the data is $\Delta
p(\theta)=\sqrt{\bar p_\theta-\bar p^2_\theta}/\sqrt{N}$.  Then, from
error propagation theory, the error on each of the angles $\theta$
estimated from $\bar p_\theta$ is
\begin{eqnarray}
  \Delta\theta={\Delta\bar
    p}/{\left|\frac{\partial
        p(\theta)}{\partial\theta}\right|}=
\sqrt{\frac{\bar p_\theta(1-\bar p_\theta)}{p(\theta)[1-p(\theta)]}}
\frac1{\sqrt N}\simeq \frac1{\sqrt N}
  \labell{eq}\;,
\end{eqnarray}
since $\bar p_\theta\to p(\theta)$ for large $N$. Thus, Bob's overall
error on his estimate of the direction of Alice's $z$ axis using $3N$
spins will be $3/\sqrt{N}$.  Notice that, 
without using entangled states at the preparation stage, the
$1/\sqrt{N}$ asymptotic scaling cannot be beaten~\cite{qmetro}.

To send a spatial direction, the exchanged qubits must possess
directional information, as is the case of spin $1/2$ particles.
Nevertheless, a partial directional information can be encoded also in
the polarization of a photon. In fact, using single photons one can
transmit a direction in the plane orthogonal to the wave vector. In
this case Bob just needs to orient his polarizers in two directions
(say the $x$ and $y$ axes), and the above procedure gives him Alice's
secret direction with accuracy $2 \sqrt{2/N}$.

{\em Optimal protocol using secret random bits.}---The previous
protocol requires $N$ qubits and $N$ secret bits to communicate a
secret spatial direction with an rms error $3\sqrt{3/N}$. Now we show
that a suitable use of entanglement allows to transmit a whole frame
of Cartesian axes with rms error $1/N$, with only $3 \log N$ secret
bits needed. To this purpose, we first recall the basic features of
the optimal protocol to publicly transmit a Cartesian
frame~\cite{refframe}, then showing how to modify it in order to
achieve unconditional security.

The optimal state for the reference frame communication requires the
decomposition of the Hilbert space $\spc H^{\otimes N}$ of $N$ spin
1/2 particles into irreducible representations of the rotation group.
This decomposition is given by
\begin{equation}\label{SpaceDecomp}
\spc H^{\otimes N}= \bigoplus_{j =0(\frac 1 2)}^{N/2}~ \spc H_j
\otimes \Cmplx^{m_j}~,
\end{equation}
where $j$ is the quantum number of the total angular momentum, ranging
from $0(\frac 1 2)$ to $N/2$ for $N$ even (odd), $\spc H_j$ is a $2j +1$
dimensional space supporting an irrep of the rotation group, and $\Cmplx^{m_j}$ is a
\emph{multiplicity space}, whose dimension $m_j$ is equal to the
number of equivalent irreps corresponding to the quantum number
$j$.  In this decomposition of the Hilbert space, the action of
a collective rotation $U_g^{\otimes N},~ g \in \SU 2$ becomes
\begin{equation}\label{Wedderburn}
U_g^{\otimes N} =  \bigoplus_{j=0(\frac 1 2)}^{N/2}~ U_g^j \otimes \openone_{m_{j}}~,
\end{equation}
where $\{U_g^j\}$ is the irreducible representation with angular
momentum $j$, and $\openone_d$ denotes the identity in a
$d$-dimensional Hilbert space. From Eq.~(\ref{Wedderburn}) it is clear
that the multiplicity spaces are the rotationally invariant subsystems
of the global Hilbert space $\spc H^{\otimes N}$.
        
For large $N$, the optimal states for the transmission of a reference
frame are given by~\cite{refframe}
\begin{equation}\label{OptState}
|A\> = \bigoplus_{j =0(\frac 1 2)}^{ \frac N 2-1} \frac{A_j}{\sqrt{2j+1}} ~|E_j\>~,
\end{equation} 
$A_j$ being suitable coefficients ($A_j \approx \sqrt{4/N} \sin
\left(\frac{2 \pi j} {N}\right)$ for $N\gg 1$), and $|E_j\> \in \spc H_j
\otimes \Cmplx^{m_j}$ being the maximally entangled vector
\begin{equation}
|E_j\>= \sum_{m=-j}^j~|jm\>|m\>~,
\end{equation}
where $\{|jm\> \in \spc H_j\}$ are the eigenstates of the total
angular momentum $J_z$, and the vector $|m\>$ runs on the first $2j+1$
elements of a basis of the multiplicity space $\Cmplx^{m_j}$. Since
the state $|A\>$ is prepared by Alice referring to \emph{her}
Cartesian frame, from Bob's point of view all spins are rotated by the
unknown rotation $g \in \SU 2$ that connects his axes with Alice's.
Accordingly, Bob receives the state $U_g^{\otimes N} |A\>$, and his
aim is to perform the best possible measurement to infer $g$.  For a
state of the form \eqref{OptState} such measurement is given by the
POVM $M(h) \d h=U_h^{\otimes N} |B\>\<B| U_h^{\dag \otimes N} \d h$,
where $\d h$ is the invariant measure over the rotation group, and
$|B\>$ is 
\begin{equation}
|B\> = \bigoplus_{j} \sqrt{2 j+1}~ |E_j\>~.
\label{eccoqui}
\end{equation}   
The use of the state $|A\>$ and of the measurement $M(h)$ allows to
communicate optimally a Cartesian frame with an asymptotic rms error
$1/N$~\cite{refframe}.  The optimality proof for this protocol is
given in Ref.~\cite{EntEstimation}. A transmission scheme using only
the state $|A\>$ is not secret, since anybody can intercept the spins,
perform the optimal measurement, and reprepare the state according to
the outcome.

To construct a secret protocol, notice that the vector $|E_j\>$ in the
state $|A\>$ can be replaced by any other maximally entangled vector
$|W_j\> = (W_j \otimes \openone_{m_j}) |E_j\>$, where $W_j$ is a local
unitary on $\spc H_j$. With the same substitution in the
POVM~(\ref{eccoqui}), the outcome probabilities in the orientation
measurement are unchanged. Thus the estimation is still optimal. The
idea then is to randomize the choice of the maximally entangled vector
$|W_j\>$ in order to make it impossible for Eve to extract any kind of
information about Alice's axes. To this purpose, consider the
unitaries $W_{j,p_j,q_j}$ defined
\begin{equation}
  W_{j,p_j,q_j} =  \sum_{m=-j}^j 
\exp\Big({\frac{2\pi i\:m q_j}{2j+1}}\Big)
  |m \oplus p_j\>\<m|~,
\end{equation}
$\oplus$ here denoting addition modulo $2j+1$. These unitaries form a
representation of the ``shift and multiply'' group $\mathbb Z^{2j+1}
\times \mathbb Z^{2j +1}$, which is irreducible on $\spc H_j$. A
completely secure communication can be obtained if Alice sends one of
the states
\begin{equation}
|A_{\{p_j, q_j\}}\> = \bigoplus_{j =0(\frac 1 2)}^{\frac N2-1}
\frac{A_j}{\sqrt{2j+1}} ~\left( W_{j, p_j,q_j} \otimes \openone_{m_{j}} \right)|E_j\>~,
\end{equation} 
chosen according to a secret random sequence $\{p_j,q_j\}$ that
she shares only with Bob. The number of possible sequences is $C=
\sum_{j=0(\frac 1 2)}^{N/2}~(2j +1)^2 \simeq {\mathcal O}(N^3)$. This
means that Alice and Bob asymptotically need $3 \log N$ bits of shared
randomness.  From the point of view of Eve, the randomization
procedure is equivalent to the preparation of the mixed state $\rho_E
= \sum_{\{p_j, q_j\}} |A_{\{p_j,q_j\}}\>\< A_{\{p_j,q_j\}}|/C$. Due to
the irreducibility of the representations $\{W_{j,p_j,q_j}\}$ this averaged
state can be easily calculated, obtaining
\begin{equation}\label{RhoEve}
\rho_E = \bigoplus_{j =0(\frac 1 2)}^{\frac N2-1} 
\frac{|A_j|^2}{(2j+1)^2}\; \openone_{2j+1} \otimes
\sum_{m=-j}^j|m\rangle\langle m|~.
\end{equation} 
Using Eq.~\eqref{Wedderburn}, it is immediate to see that Eve's state
$\rho_E$ is completely invariant under rotations, therefore no useful
information can be extracted from it about the orientation of Alice's
axes.  On the other hand, since Bob knows which pure state
$|A_{\{p_j,q_j\}}\>$ was sent, he can perform the optimal orientation
measurement. He then recovers Alice's axes with the asymptotically
optimal rms error $1/N$.

This result sheds a new light on the role of private shared reference
frames (SRF) as a physical resource. In fact, we can relate the above
protocol with the cryptographic protocol of Ref.~\cite{qcrypt}, where
a private SRF is used to communicate a secret string of bits. While in
our case we have asymptotically
\begin{equation}
N~ qbits + 3 \log N~ secret~ bits \longrightarrow  private~SRF~,
\end{equation}
in the case of Ref.~\cite{qcrypt} one has
\begin{equation}
N~ qbits +  private~ SRF \longrightarrow 3 \log N~ secret~ bits~.
\end{equation} 
In other words, the comparison of the two results gives a tight
balance between the number of secret random bits needed to establish a
private SRF and the secret classical capacity associated to it.

Is it really necessary to have a string of secret bits to establish a
private SRF? Remarkably, this is not the case. In the following we
sketch two alternative protocols, modeled on the BB84~\cite{bb84} and
the Ekert protocols~\cite{ekert}, in which the private SRF is
established by only using quantum and classical communication over an
authenticated public channel:
The main idea of both protocols is to exploit the fact that the
information encoded in the multiplicity spaces $\Cmplx^{m_j}$ of
Eq.~\eqref{SpaceDecomp} is frame-independent~\cite{prl}, since
according to Eq.~\eqref{Wedderburn} the multiplicity spaces are
invariant under rotations. Thus, even though Alice and Bob do not
share a Cartesian frame, they can exchange qubits and test whether an
eavesdropper is acting in the rotationally invariant subsystems of the
Hilbert space. If the security level is too low, they can decide to
abort the protocol (before any reference information is transmitted).
This strategy prevents Eve to access the information encoded in the
multiplicity spaces, namely her POVM must have the form $P_i =
\bigoplus_j P_{ij} \otimes \openone_{m_j}$, for some suitable $P_{ij}
\ge 0$, $\sum_i P_{ij}= \openone_{2j+1}$.  Since for states of the
form \eqref{OptState} representation spaces and multiplicity spaces
are maximally entangled, Eve's measurement gives no information about
Alice's axes (again the state seen by Eve is the $\rho_E$ of
Eq.~\eqref{RhoEve}).

{\em BB84-type protocol}---With probability $1/2$ Alice
sends the state $U_h^{\otimes N} |A\>$, where $h$ is a random
rotation, otherwise she sends a test-state.
The possible test states are 
\begin{equation}
\tau_{j,m}= \frac{\openone_{2j+1}}{2j+1} \otimes |m\>\<m|
\mbox{ and }
\tilde \tau_{j,m}= \frac{\openone_{2j+1}}{2j+1} \otimes |\tilde m\>\<\tilde m|~,
\end{equation}
where $\{|m\>\}$ and $\{|\tilde m\>\}$ are two bases of $\Cmplx^{m_j}$
related by a Fourier transform. An eavesdropper cannot tell whether
the state $U_h^{\otimes N} |A\>$ or the test states were sent, as
there is a large overlap between them, i.e. a fidelity
$F=A_j^2/(2j+1)^2$ (of order $1/N^3$ for $j\approx N/4$). On the other
hand, Bob performs with probability $1/2$ the optimal measurement of
orientation, otherwise he performs a test-measurement. For the
test-measurement he randomly chooses one of the two von Neumann
measurement $V_{j,m}= \openone_{2j+1} \otimes |m\>\<m|$, or
$\widetilde V_{j,m}=\openone_{2j+1} \otimes |\tilde m\>\<\tilde m|$.
Then, as in the BB84 protocol, using the authenticated public
classical channel, Alice and Bob declare the ``basis'' they used:
Alice announces whether she prepared an orientation-state, a
test-state of the kind $\tau$, or a test state of the kind $\tilde
\tau$, and similarly Bob announces whether he measured the
orientation, the observable $V$, or the observable $\widetilde V$.
They keep only the cases where Bob's measurement coincided with
Alice's preparation, and discard the rest.  Alice announces the values
of $j$ and $m$ for the test states she sent, so that Bob can check
whether or not the results of his measurements match with her data. In
this way, an eavesdropper in the rotationally invariant subsystems can
be detected, and Bob can decide to abort the protocol, if the security
level is not sufficiently high.  Otherwise, Alice publicly
communicates the random rotation $h$ she performed on the optimal
state $|A\>$. On each of the states Alice sent, Bob estimated the
rotation $gh$ with rms error $1/N$, where $g$ is the unknown rotation
connecting his axes with Alice's ones. Thus, knowing $h$, Bob can
immediately infer $g$ and align his axes. Notice that here $N$ is the
number of spins in each of the states that Alice sends. If she sends
$M$ states, the total number of transmitted spins is $MN$ and Bob's
accuracy will scale as $4/\sqrt{M}N$ (where the square root comes from
the central limit theorem, and the factor $4$ refers to the fact that
only $1/4$ of the states are used in average for the orientation
measurement). Compared with the previous protocol, the scaling in
accuracy with the total number $MN$ of spins is sub-optimal, but the
decrease in accuracy is needed because of the elimination of the prior
shared random bits.

{\em Ekert-type protocol.}---Alice starts with two sets of $N$ spin 1/2
particles, half of which she sends to Bob. The Hilbert spaces of both
sets of spins, $\spc H_A^{\otimes N}$ and $\spc H_B^{\otimes N}$, can
be decomposed as in Eq.~(\ref{SpaceDecomp}), and she can prepare the
entangled state
\begin{equation}
|\Phi\>_{AB} = \bigoplus_{j=0(\frac 1 2)}^{\frac N 2 -1}~ \frac{A_j}{\sqrt{(2j+1)~m_j}}
\sum_{m,n}~|jm\>_A |n\>_A |jm\>_B |n\>_B~,
\end{equation}  
where $A_j$ are the same coefficients as in the optimal state in
Eq.~(\ref{OptState}). Note that for any value of the angular momentum
the state $|\Phi\>_{AB}$ exhibits a maximal entanglement between the
multiplicity spaces $\Cmplx^{m_j}_A$ and $\Cmplx_B^{m_j}$. For this
reason, an eavesdropper in the rotationally invariant subsystems can
be detected by Alice and Bob by testing Bell inequalities, as in the
original Ekert protocol: to evade detection Eve must not act on the
rotationally invariant subsystems of Bob.  A SRF is established when
Alice and Bob both perform the optimal orientation-measurement, given
by Eq.~\eqref{eccoqui}. In fact, the probability density that they
measure the rotations $h_A$ and $h_B$ is $p(h_A, h_B) =\< \Phi|~M(h_A)
\otimes {U_g^{\otimes N}}^\dag M(h_B) U_g^{\otimes N} |\Phi\>$, where
$g$ is the unknown rotation connecting their axes. A short calculation
allows to derive the following equality
\begin{eqnarray}
p(h_A, h_B) =\< \Phi|~M(h_A) \otimes {U_g^{\otimes N}}^\dag M(h_B)  
U_g^{\otimes N} |\Phi\>&&\\
\nonumber
= \<A| M(h_A h_B^{-1} g)|A\>~, \qquad\qquad\qquad\qquad\qquad&&
\end{eqnarray} 
which implies that $h_A h_B^{-1}g$ is distributed with the rms error
$1/N$ of the optimal orientation measurement, even though the outcomes
$h_A$ and $h_B$ are completely random, as the local states of Alice
and Bob are rotationally invariant. Once Alice has communicated (with
the authenticated public channel) the outcome $h_A$ of her
measurement, Bob can use this information to retrieve the unknown
rotation $g$ with the precision $1/N$.  Again, since Eve cannot touch
the rotationally invariant subsystems, she can gain no information
about Alice's axes.  Also here, part of the exchanged spins is
employed (in the Bell inequalities tests) to make up for the absence
of the prior shared secret bits.

{\em Conclusions.}--- In this Letter we have shown that quantum
mechanics allows one to secretly communicate directions in space,
either with or without the need of prior secret random bits. We have
given a simple protocol that needs no entanglement and an entangled
protocol that achieves the ultimate bounds in the precision of
reference frame transmission. The unentangled protocol can be easily
implemented experimentally using spin 1/2 particles or single photon
polarization states. Two further protocols that do not need secret
random bits have also been presented.  The ideas of this paper have
been presented for spatial reference frames, however, they can be
exploited to achieve the secret transmission of any possible reference
frame, e.g.  to obtain a secret clock synchronization.

\begin{acknowledgments}
  Financial support comes from MIUR through PRIN 2005 and from the EU
  through the project SECOQC (IST-2003-506813).
\end{acknowledgments}

\end{document}